\documentclass[prd,superscriptaddress,nofootinbib,tightenlines]{revtex4}
\usepackage{epsfig}
\usepackage[colorlinks=true,linkcolor=blue,urlcolor=blue,citecolor=blue]{hyperref}
\usepackage{amsmath}
\usepackage{amsfonts}
\usepackage{amssymb}
\usepackage{bm}

\newcommand{\ben}{\begin{displaymath}}
\newcommand{\een}{\end{displaymath}}
\newcommand{\be}{\begin{equation}}
\newcommand{\ee}{\end{equation}}
\newcommand{\bea}{\begin{eqnarray}}
\newcommand{\eea}{\end{eqnarray}}

\newcommand{\nn}{\nonumber \\ }

\usepackage{comment}
\begin{document}
\title{Local spatial densities for composite spin-3/2 systems}
 \author{H.~Alharazin}
  \affiliation{Institut f\"ur Theoretische Physik II, Ruhr-Universit\"at Bochum,  D-44780 Bochum,
 Germany}
\author{B.-D.~Sun}
  \affiliation{Guangdong Provincial Key Laboratory of Nuclear Science,
Institute of Quantum Matter, \\South China Normal University, Guangzhou 510006, China} 
\affiliation{Guangdong-Hong Kong Joint Laboratory of Quantum Matter,\\
Southern Nuclear Science Computing Center, \\South China Normal University, Guangzhou 510006, China
}
\affiliation{Helmholtz-Institut f\"ur Strahlen- und Kernphysik and Bethe
  Center for Theoretical Physics, Universit\"at Bonn, D-53115 Bonn, Germany}
\author{E.~Epelbaum}
 \affiliation{Institut f\"ur Theoretische Physik II, Ruhr-Universit\"at Bochum,  D-44780 Bochum,
 Germany}
\author{J.~Gegelia}
 \affiliation{Institut f\"ur Theoretische Physik II, Ruhr-Universit\"at Bochum,  D-44780 Bochum,
 Germany}
\affiliation{Tbilisi State  University,  0186 Tbilisi,
 Georgia}
 \author{U.-G.~Mei\ss ner}
 \affiliation{Helmholtz-Institut f\"ur Strahlen- und Kernphysik and Bethe
   Center for Theoretical Physics, Universit\"at Bonn, D-53115 Bonn, Germany}
 \affiliation{Institute for Advanced Simulation, Institut f\"ur Kernphysik
   and J\"ulich Center for Hadron Physics, Forschungszentrum J\"ulich, D-52425 J\"ulich,
Germany}
\affiliation{Tbilisi State  University,  0186 Tbilisi,
 Georgia}

\date{22 December 2022}
\begin{abstract}
The definition of local spatial densities by using sharply localized one-particle states is
applied to spin-3/2 systems. 
Matrix elements of the electromagnetic current and the energy-momentum tensor are considered and 
integral expressions of associated spatial distributions in terms of form factors are derived. 

\end{abstract}

\maketitle

\section{Introduction}

In close analogy with the electric charge density
of hadrons \cite{Hofstadter:1958,Ernst:1960zza,Sachs:1962zzc}
it has been suggested to interpret the Fourier transforms of the
gravitational form factors in the Breit frame as local densities of physical quantities
characterizing various composite systems \cite{Polyakov:2002wz,Polyakov:2002yz,Polyakov:2018zvc}.
The identification of the spatial densities  with the Fourier
transforms of the electromagnetic and gravitational form factors in the Breit frame, especially
for systems with intrinsic sizes comparable to their Compton wavelengths,  has been questioned in
Refs.~\cite{Burkardt:2000za,Miller:2007uy,Miller:2009qu,Miller:2010nz,Jaffe:2020ebz,Miller:2018ybm,Freese:2021czn}.   
This issue has raised much interest recently
\cite{Lorce:2020onh,Lorce:2022cle,Lorce:2018egm,Chen:2022smg,Guo:2021aik,Panteleeva:2021iip,Epelbaum:2022fjc,Panteleeva:2022khw,Kim:2021kum,Kim:2021jjf,Kim:2022bia,Kim:2022wkc,Freese:2021mzg,Freese:2022fat,Carlson:2022eps}.
On the one hand, the formalism of Wigner phase space distributions is utilized in
Ref.~\cite{Lorce:2018egm} and in subsequent publications. 
On the other hand, two-dimensional densities in the transverse plane obtained in the formalism
of light front dynamics by integrating over the $x^-$ 
coordinate have been considered as the only possible true internal
densities \cite{Freese:2021czn,Freese:2022fat}. Note that two-dimensional densities have been also 
considered earlier in Ref.~\cite{Lorce:2018egm}.
A definition of spatial densities of local operators using sharply localized wave packet
states, applicable to systems with arbitrary Compton wavelengths, 
has been suggested in Ref.~\cite{Epelbaum:2022fjc}, see also Ref.~\cite{Fleming:1974af} for
a related earlier study.  Specifying the one-particle state of a spin-0 system by a
spherically symmetric wave packet localized in space and taking the size of the packet
much smaller than all internal characteristic scales of the considered system,
spatial charge distributions have been defined in the zero average momentum frame (ZAMF).
The new definition has been also generalized to moving Lorentz frames. 
Recently this definition has been also applied to electromagnetic densities of spin-1/2
systems \cite{Panteleeva:2022khw}, and to gravitational densities of spin-0 and spin-1/2 systems,
defined via the matrix elements of the energy-momentum tensor (EMT)  \cite{Panteleeva:2022uii}. 

In the current work we consider spatial densities corresponding to the electromagnetic current
and the EMT for spin-3/2 systems.  We obtain the corresponding expressions in terms of
form factors and discuss their physical interpretation. 

Our work is organized as follows. In Sec.~\ref{EmD} we define spatial densities corresponding
to the electromagnetic current for spin-3/2 systems. 
Section~\ref{GrD} deals with the one-particle matrix elements of the EMT.
In Sec.~\ref{LDB} we discuss the large-distance behavior of various 
distributions of the delta resonance, which is the most studied composite
spin-3/2 system, in chiral EFT. 
We summarize our results in Sec.~\ref{summary}.
The appendix contains the  lengthy expressions of various quantities. 

\section{Electromagnetic densities in the zero average momentum frame}
\label{EmD}

We choose the four-momentum eigenstates
$|p,s\rangle$ characterizing our spin-3/2 system to be normalized as
\begin{equation}
\langle p_f,s'|p_i,s\rangle = 2 E (2\pi )^3 \delta_{s's}\delta^{(3)} ({\bf p}_f-{\bf p}_i)\,,
\label{NormStateN}
\end{equation}
where $(p_i,s)$ and $(p_f,s')$ are the momentum and polarization of the initial and final state,
respectively. Further, $p=(E,{\bf p})$ with $E=\sqrt{m^2+{\bf p}^2}$, where $m$ is the particle's mass. 

To define the spatial densities via the matrix elements of local operators we use normalizable
Heisenberg-picture states written in terms of wave packets as follows:
\begin{equation}
  |\Phi, {\bf X},s \rangle = \int \frac{d^3 {p}}{\sqrt{2 E (2\pi)^3}}  \, \phi(s,{\bf p})
  \, e^{-i {\bf p}\cdot{\bf X}} |p ,s \rangle,  
\label{statedefN2}
\end{equation}
where the parameters ${\bf X}$ are interpreted as the coordinates of the center of the
charge or mass distribution, corresponding to the operator under consideration,
and the profile function satisfies the normalization condition
\begin{equation}
\int d^3 {p} \,  | \phi(s,{\bf p})|^2 =1\,.  
\label{normN}
\end{equation}
To {\it define} the density distributions of the system we use spherically symmetric wave
packets and profile functions of which $\phi(s, {\bf p}) = \phi({\bf p}) = \phi(|{\bf p} |)$
are also spin-independent.  
The average of the three-momentum of the system vanishes in states corresponding to such
packets, thus they describe the system in the ZAMF.
For our calculations it is convenient to define dimensionless profile functions 
\begin{equation}
\phi({\bf p}) = R^{3/2} \, \tilde \phi(R  {\bf p})  \,,
\label{packageFormN}
\end{equation} 
where $R$ specifies the size of the wave packet. Small values of $R$ correspond to sharp
localization of the packet.

The matrix elements of the electromagnetic current operator between momentum eigenstates of a
spin-3/2 system can be parameterized in terms of four form factors, 
see Ref.~\cite{Pascalutsa:2006up} for a review. We use here the notation of Ref.~\cite{Cotogno:2019vjb}:
\begin{eqnarray}
\langle p_f, s'| J_{\mu}| p_i,s \rangle &=& - \bar u^\beta (p_f,s') \Biggl[ \frac{P_\mu }{m}  \left( g_{\alpha\beta} F_{1,0}^V (q^2) -\frac{q_\alpha q_\beta }{2 m^2} \, F_{1,1}^V(q^2)\right) \nonumber\\
&+& \frac{i}{2m}  \sigma_{\mu\rho}  q^\rho \left( g_{\alpha\beta} F^V_{2,0} (q^2) -\frac{q_\alpha q_\beta }{2 m^2} \, F^V_{2,1}(q^2)\right)
\Biggr]  u^\alpha(p_i,s) \,,
\label{defFFsC}
\end{eqnarray}
where $P=(p_i+p_f)/2$, $q=p_f-p_i$.  
In terms of these variables, the energies are given as $E=(m^2+ {\bf P}^2 - {\bf P}\cdot
{\bf q} +{\bf q}^2/4)^{1/2} $ and $E'=(m^2+ {\bf P}^2 + {\bf P}\cdot {\bf q} +{\bf q}^2/4)^{1/2} $.
The spinors of the spin-3/2 states are defined as follows:
\begin{eqnarray}
  u^\mu(p,s) & = & \sum_{\lambda, \sigma} \langle 1\lambda, \frac{1}{2}\sigma|\frac{3}{2} s\rangle
  e^\mu(p,\lambda)  u(p,\sigma)\,,\nonumber\\
e^\mu (p,\lambda) &=& \left( \frac{ \hat {\bf e}_\lambda\cdot {\bf p}}{m}, \hat {\bf e}_\lambda+ \frac{ {\bf p} ( \hat {\bf e}_\lambda\cdot {\bf p})}{m(p_0+m)}  \right)\,, \nonumber\\
u(p,\sigma) & = &  \sqrt{p_0+m}  \left( \chi_\sigma,  \frac{ {\bm \sigma}\cdot{\bf p} }{p_0+m} 
\, \chi_\sigma\right)^T\,,
\label{defspinors}
\end{eqnarray}
where  $ \langle 1\lambda, \frac{1}{2}\sigma|\frac{3}{2} s\rangle$
are the pertinent Clebsh-Gordon coefficients and
\begin{equation}
\hat{\bf e}_+ = -\frac{1}{\sqrt{2}} \left(1,i,0 \right) , \qquad  \hat{\bf e}_0 = \left(0,0,1 \right) , \qquad  \hat{\bf e}_- = \frac{1}{\sqrt{2}} \left(1,-i,0 \right) . 
\label{defevec}
\end{equation}
The Dirac spinors are normalized as $\bar u(p,s') u(p,s) = 2 m\, \delta_{s's}$.
The matrix element of the electromagnetic current operator in localized states is given by 
\begin{eqnarray}
 j^{\mu}_{\phi}(s',s,{\bf r})  &\equiv &  \langle \Phi, {\bf X},s' | \hat
                                   J^{\mu} ({\bf x}, 0 ) | \Phi, {\bf X},s
                                   \rangle 
                                    \nn &
  = & -   \int \frac{d^3 {P} \, d^3 {q}}{(2\pi)^3 \sqrt{4 E
    E'}}\, 
 \bar u^\beta \left( P+\frac{q}{2},\sigma' \right) \Biggl[  \frac{P_\mu }{m}  \left( g_{\alpha\beta} F_{1,0}^{V} (q^2) -\frac{q_\alpha q_\beta }{2 m^2} \, F_{1,1}^{V}(q^2)\right) \nonumber\\
&+& \frac{i}{2m}  \sigma_{\mu\rho}  q^\rho \left( g_{\alpha\beta} F^{V}_{2,0} (q^2) -\frac{q_\alpha q_\beta }{2 m^2} \, F^{V}_{2,1}(q^2)\right) \Biggr]  
u^\alpha\left(P-\frac{q}{2},\sigma\right)  \phi\bigg({\bf P} -
\frac{\bf q}{2}\bigg)  \phi^\star\bigg({\bf P} +\frac{\bf q}{2}\bigg)  e^{ - i {\bf q}\cdot {\bf  r}} ,
\label{EMC}
\end{eqnarray}
where ${\bf r} = {\bf x} - {\bf X}$. 
By applying the method of dimensional counting of Ref.~\cite{Gegelia:1994zz}, the leading
contributions in Eq.~(\ref{EMC}) for $R\to 0$ can be obtained without 
specifying the form of the form-factors and the profile function $\phi(|\bf{p}|)$.
Provided that the form factors $F_{1,0}^{V} (q^2)$, $F_{1,1}^{V} (q^2)$, $F_{2,0}^{V} (q^2)$ and 
$F_{2,1}^{V} (q^2)$ decay for large $q^2$ as $1/q^2$, $1/q^4$, $1/q^3$, $1/q^5$ (or faster), respectively,
the only non-vanishing contribution for $R\to 0$ is generated from the region of integration
where ${\bf P}$ is large. It can be obtained by substituting ${\bf P}=  {\bf Q}/R$, expanding
the resulting integrand in Eq.~(\ref{EMC}) in powers of $R$
around $R=0$ and keeping the leading order term for each component separately. Introducing
$\hat n=  {\bf Q}/|{\bf Q}| $ and using the spherical symmetry of the wave packet, the 
integration over $|{\bf Q}|$ can be carried out without specifying the radial profile function.
The result of the integration is given below in terms of irreducible tensors and the multipole
operators which are defined as follows. The $n$-th rank irreducible tensors in coordinate and
momentum spaces, respectively, are given by ($r\neq 0 $)
\begin{equation}
Y_n^{i_1i_2\cdots i_n}(\Omega_r) = \frac{(-1)^n}{(2n-1)!!} \, r^{n+1} \partial^{i_1} \partial^{i_2} \cdots \partial^{i_n} \frac{1}{r} \,, \qquad 
Y_n^{i_1i_2\cdots i_n}(\Omega_p) = \frac{(-1)^n}{(2n-1)!!} \, p^{n+1} \partial^{i_1} \partial^{i_2} \cdots \partial^{i_n} \frac{1}{p} \,.
\label{defIrrTens}
\end{equation} 
The quadrupole- and octupole-operators, $\hat Q^{ij}$ and $\hat O^{ijk}$, for a spin-3/2 system
are given in terms of the spin operator $\hat S^i$ via  
\begin{eqnarray}
\hat Q^{ij} & = & \frac{1}{2} \left( \hat S^i\hat S^j+\hat S^j\hat S^i-\frac{2}{3} S(S+1) \delta^{ij}\right) \,, \nonumber\\
\hat O^{ijk} & = & \frac{1}{6} \left( \hat S^i \hat S^j \hat S^k +\hat S^j\hat S^i \hat S^k + \hat S^k \hat S^j \hat S^i 
 +\hat S^j\hat S^k \hat S^i+ \hat S^i \hat S^k \hat S^j +\hat S^k\hat S^i \hat S^j \right. \nonumber\\
 & - & \left. \frac{6 S(S+1)-2}{5} \left(\delta^{ij} \hat S^k + \delta^{ik} \hat S^j +\delta^{kj} \hat S^i  \right)\right) \,,
\label{defQs}
\end{eqnarray} 
with $i,j,k=1,2,3$.

The results for the matrix elements after integration over $|{\bf Q}|$ in Eq.~(\ref{EMC})
have the form:
\begin{eqnarray}
j^{0}_{\phi}(s',s,{\bf r})   &=& \int \frac{d^3 {q}}{(2\pi)^3 }\, e^{ - i {\bf q}\cdot {\bf  r}} \frac{1}{4 \pi} \int d^2 \hat{n} \left\{   \mathcal{Z}_0(-q_\perp^2)~\delta_{s' s} + \left[ \mathcal{Z}_1(-q_\perp^2) ~ \hat n^k\hat n^l 
 +  \mathcal{Z}_2 (-q_\perp^2)~  \frac{q^k_\perp q^l_\perp}{{m^2}}  \right]  {\hat Q^{k l}_{s' s}} \right\}\, 
 \\
&=&  \rho^C_{0}\left(r\right)\delta_{s's}+\rho^C_{2}\left(r\right)Y_{2}^{kl}\left(\Omega_{r}\right)\hat{Q}_{s's}^{kl},
\label{j0}\\
j^{i}_{\phi}(s',s,{\bf r})   &=& \int \frac{ d^3 {q}}{(2\pi)^3 }\, e^{ - i {\bf q}\cdot {\bf  r}} \frac{i}{4 \pi} \int d^2 \hat{n}  \left\{  \left[ \mathcal{A}_0(-q_\perp^2) ~\hat n^i \hat n^l  \epsilon^{k l n } 
+  \mathcal{A}_1(-q_\perp^2)  \left( ~ \delta^{ k l } -  \hat n^k  \hat n^l \right)  \epsilon^{i l n  }\right] \frac{q_\perp^n}{m} {\hat S^{ k }_{s' s}}  \right.
\nn
&+&
\bigg[ \left(  \mathcal{A}_2(-q_\perp^2)~ \hat n^t \hat n^z +  \mathcal{A}_3(-q_\perp^2)~ \frac{q_\perp^t q_\perp^z}{{m^2}}  \right) \hat n^i  \hat n^l \epsilon^{k l n }\nn &+&     \left(  \mathcal{A}_4(-q_\perp^2)\hat n^t \hat n^z + \mathcal{A}_5(-q_\perp^2)~ \frac{q_\perp^t q_\perp^z}{{m^2}} \right)  \epsilon^{i l n}   \left( \delta^{ k l} - \hat n^k \hat n^l  \right) \bigg] \frac{q_\perp^n}{m} {\hat {O}^{ k t z }_{s' s}} \Biggl\}\,\\
&=& { i 
\epsilon^{ikn}\hat{S}_{s's}^{k}Y_{1}^{n} \frac1m\frac{d}{dr} \rho^M_{1} \left(r\right) \text{ }+ i  \epsilon^{ikn}\hat{O}_{s's}^{ktz}Y_{3}^{ntz}\, \frac{r^{3}}{m^3}\left(\frac{1}{r}\frac{d}{dr}\right)^{3} \rho^M_{3}\left(r\right),
}
\label{ji}
\end{eqnarray}
where $q_\perp^2={\bf q}^2 - ({\bf q}\cdot \hat n)^2$
and the coefficient functions $ \mathcal{Z}_i$ and $ \mathcal{A}_i$ are given in the appendix.
The quantities  $ \rho^C_{0}(r)$ and $\rho^C_{2}(r)$ are the monopole and quadrupole charge
densities, respectively, whereas $\rho^M_1(r) $ and  $\rho^M_3(r)$ are the dipole and
octupole scalar magnetization densities, respectively. These quantities are given by
\begin{eqnarray}
  \rho^C_{0} (r) &=& \frac{1}{4\pi}\int\frac{d^{3}q}{(2\pi)^{3}}\,e^{-i{\bf q}\cdot{\bf r}} \int d^{2}\hat{n} \Biggl\{ F_{1,0}^{V}(-q_\perp^2)  + \frac{q_\perp^2}{6 m^2} \left[ -2 F_{1,0}^{V}(-q_\perp^2) +  F_{1,1}^{V}(-q_\perp^2) + F_{2,0}^{V}(-q_\perp^2) \right] \nn &+& \frac{q_\perp^4}{24m^4} \left[ -2 F_{1,1}^{V}(-q_\perp^2) + F_{2,1}^{V}(-q_\perp^2) \right] \Biggl\}~
 ,\\
 \rho^C_{2} (r) &=&  -\frac{r}{4\pi}\frac{d}{dr}\frac{1}{r}\frac{d}{dr} \int\frac{d^{3}q}{(2\pi)^{3}}\,e^{-i{\bf q}\cdot{\bf r}} \int d^{2}\hat{n} \,  \frac{q_\perp^2}{12 m^2 q^2} \Biggl\{  4 F_{1,0}^{V}(-q_\perp^2) + F_{1,1}^{V}(-q_\perp^2) - 2 F_{2,0}^{V}(-q_\perp^2) \nn &-&   \frac{3 q_\perp^2}{ q^2} \left[   F_{1,0}^{V}(-q_\perp^2)  + F_{1,1}^{V}(-q_\perp^2)  \right] -   \frac{q_\perp^2}{2 m^2} \left[ - 2 F_{1,1}^{V}(-q_\perp^2)  + F_{2,1}^{V}(-q_\perp^2) \right]  -  \frac{3q_\perp^4}{4 m^2 q^2} F_{1,1}^{V}(-q_\perp^2)\Biggl\},
\label{j0_Ztilde}
\end{eqnarray}
\begin{eqnarray}
\rho^M_1(r) &=& \frac{1}{4\pi } \int\frac{d^{3}q}{(2\pi)^{3}}\,e^{-i{\bf q}\cdot{\bf r}} \int d^{2}\hat{n} \frac{q_{\perp}^{2}}{2q^{2}} \bigg\{ F_{1,0}^{V}(-q_\perp^2) -\frac{1}{3}F_{2,0}^{V}(-q_\perp^2) 
\nn
&+&  \frac{q_{\perp}^{2}}{30 m^2} \left[ -2 F_{1,0}^{V}(-q_\perp^2) + 7  F_{1,1}^{V}(-q_\perp^2) \right. + \left. 2  F_{2,0}^{V}(-q_\perp^2) -2  F_{2,1}^{V}(-q_\perp^2) \right] 
\nn
&+& \frac{q_{\perp}^{4}}{60 m^4} \left[ F_{2,1}^{V}(-q_\perp^2)  - F_{1,1}^{V}(-q_\perp^2)  \right]  \bigg\},\\
\rho^M_3(r)&=& \frac{1}{4\pi} \int\frac{d^{3}q}{(2\pi)^{3}}\,e^{-i{\bf q}\cdot{\bf r}} \int d^{2}\hat{n}\frac{q_{\perp}^{4}}{48 q^4}\bigg\{ - 4 F_{1,0}^{V}(-q_\perp^2) - F_{1,1}^{V}(-q_\perp^2) + 4 F_{2,0}^{V}(-q_\perp^2) + F_{2,1}^{V}(-q_\perp^2) 
\nn &+&   \frac{5 q_{\perp}^{2}}{ q^2} \left[  F_{1,0}^{V}(-q_\perp^2) +  F_{1,1}^{V}(-q_\perp^2) - F_{2,0}^{V}(-q_\perp^2) - F_{2,1}^{V}(-q_\perp^2) \right]  +\frac{q_\perp^2 }{m^2 }\left[F_{2,1}^{V}(-q_\perp^2) -F_{1,1}^{V}(-q_\perp^2)\right]  \nn &+&  \frac{5 q_{\perp}^{4}}{4 m^2 q^2} \left[ F_{1,1}^{V}(-q_\perp^2) -  F_{2,1}^{V}(-q_\perp^2) \right] \bigg\} .
\label{J_i_Atilde}
 \end{eqnarray}

The standard expressions of the densities in terms of the form factors in the Breit frame,
$F_{i,j}(q^2)=F_{i,j}(-{\bf{q}}^2)$, which we will refer to as "naive", are obtained by first
approximating the integrand in Eq.~(\ref{EMC}) by the two leading terms in the
$1/m-$expansion\footnote{Factors of $m$ introduced in Eqs.~(\ref{defFFsC}) and (\ref{EMTdefJH})
for dimensional reasons in the parametrization
of the matrix elements in terms of form factors are not counted when expanding in $1/m$
neither here nor below in the case of EMT matrix elements.}
and subsequently localizing the wave packet by taking the limit $R\mapsto 0$.
The resulting expressions have the form:
\begin{eqnarray}
j^{0}_{\text{naive}}(s',s,{\bf r})  &=& \int \frac{d^3q}{(2\pi)^3} e^{ - i {\bf q}\cdot {\bf  r}} \Biggl\{ \left[  F^{V}_{1,0} \left( - {\bf q}^2 \right) +  \frac{ {\bf q}^2 }{6 m^2}F^{V}_{1,1} \left( - {\bf q}^2 \right)\right] \delta_{s' s}- F^{V}_{1,1} \left( - {\bf q}^2 \right)  \frac{q^k q^l}{6 m^2 }    {\hat Q^{ k l}_{s' s}} \Biggl\}
 ,\nn
j^{i}_{\text{naive}}(s',s,{\bf r})  & = & \int \frac{d^3q}{(2\pi)^3} e^{ - i {\bf q}\cdot {\bf  r}} i \epsilon^{i k n} \frac{q^n}{3 m} 
\Biggl\{ \left[ F^{V}_{2,0} \left( - {\bf q}^2 \right) + \frac{{\bf q}^2}{5 m^2} F^{V}_{2,1} \left( - {\bf q}^2 \right) \right]   {\hat S^{ k }_{s' s}}  
-F^{V}_{2,1} \left( - {\bf q}^2 \right)  \frac{q^l q^z}{2 m^2}   {\hat O^{ k l z }_{s' s}}  
\Biggl\} .
\end{eqnarray}

\section{Gravitational densities in the zero average momentum frame}
\label{GrD}

One is often interested in matrix elements of the quark and gluon contributions to the EMT.
As these are not separately conserved, we parameterize the matrix element of a symmetric EMT
for spin-3/2 states  in terms of ten form factors as follows \cite{Cotogno:2019vjb,Kim:2020lrs}:
\begin{eqnarray} 
\langle p_f, s'| { T_{\mu\nu} \left( \vec{x} \right) } | p_i,s \rangle &=& - \bar u^\beta (p_f,s') \Biggl[ \frac{P_\mu P_\nu}{m}  \left( g_{\alpha\beta} F_{1,0} (q^2) -\frac{q_\alpha q_\beta }{2 m^2} \, F_{1,1}(q^2)\right) \nonumber\\
&+&  \frac{q_\mu q_\nu-\eta_{\mu\nu} q^2}{4 m} \left( g_{\alpha\beta} F_{2,0} (q^2) -\frac{q_\alpha q_\beta }{2 m^2} \, F_{2,1}(q^2)\right)   \nonumber\\
&+& mg^{\mu\nu}\left(g_{\alpha'\alpha}F_{3,0}(t) - \frac{\Delta_{\alpha'} \Delta_{\alpha}}{2m^{2}}F_{3,1}(t)\right) \nonumber\\
&+& \frac{i}{2}  \frac{\left(P_\mu \sigma_{\nu\rho} + P_\nu \sigma_{\mu\rho} \right) q^\rho}{m}  \left( g_{\alpha\beta} F_{4,0} (q^2) -\frac{q_\alpha q_\beta }{2 m^2} \, F_{4,1}(q^2)\right)
 \nonumber\\ &-& 
 \frac{1}{m} \, \left(  
g_{\nu\beta} q_\mu q_\alpha + g_{\mu\beta} q_\nu q_\alpha+ g_{\nu\alpha} q_\mu q_\beta+ g_{\mu\alpha} q_\nu q_\beta -2 g_{\mu\nu} q_\alpha q_\beta   \right. \nonumber\\ &-& \left.   
g_{\mu\beta}g_{\nu\alpha} q^2 -  g_{\nu\beta}g_{\mu\alpha} q^2
\right) F_{5,0}(q^2) \nonumber\\
&+& m  (g^{\mu}_{\alpha'}g^{\nu}_{\alpha} + g^{\nu}_{\alpha'}g^{\mu}_{\alpha})F_{6,0}(t)
\Biggr]  u^\alpha(p_i,s)  e^{ - i {\bf q}\cdot {\bf  r} }  \,,
\label{EMTdefJH}
\end{eqnarray}
where in case of a conserved EMT the form factors $F_{3,0}(t)$, $F_{3,1}(t)$ and $F_{6,0}(t)$ vanish.  
The matrix element of the EMT in localized states is written as
\begin{eqnarray}
 t^{\mu\nu}_{\phi}({\bf r})  &\equiv &  \langle \Phi, {\bf X},s' | \hat
                                   T^{\mu\nu} ({\bf x}, 0 ) | \Phi, {\bf X},s
                                   \rangle 
                                    \nn &
  = & -   \int \frac{d^3 {P} \, d^3 {q}}{(2\pi)^3 \sqrt{4 E
    E'}}\, 
 \bar u^\beta \left( P+\frac{q}{2},\sigma' \right) \Biggl[ \frac{P_\mu P_\nu}{m}  \left( g_{\alpha\beta} F_{1,0} (q^2) -\frac{q_\alpha q_\beta }{2 m^2} \, F_{1,1}(q^2)\right) \nonumber\\
&+&  \frac{q_\mu q_\nu-\eta_{\mu\nu} q^2}{4 m} \left( g_{\alpha\beta} F_{2,0} (q^2) -\frac{q_\alpha q_\beta }{2 m^2} \, F_{2,1}(q^2)\right)   
+ \frac{i}{2}  \frac{\left(P_\mu \sigma_{\nu\rho} + P_\nu \sigma_{\mu\rho} \right) q^\rho}{m}  \left( g_{\alpha\beta} F_{4,0} (q^2) -\frac{q_\alpha q_\beta }{2 m^2} \, F_{4,1}(q^2)\right) \nonumber\\
&-& \frac{1}{m} \, \left(  
g_{\nu\beta} q_\mu q_\alpha + g_{\mu\beta} q_\nu q_\alpha+ g_{\nu\alpha} q_\mu q_\beta+ g_{\mu\alpha} q_\nu q_\beta -2 g_{\mu\nu} q_\alpha q_\beta   \right. \nonumber\\ &-& \left.   
 g_{\mu\beta}g_{\nu\alpha} q^2 -  g_{\nu\beta}g_{\mu\alpha} q^2
\right) F_{5,0}(q^2)  
\Biggr]  
u^\alpha\left(P-\frac{q}{2},\sigma\right)  \phi\bigg({\bf P} -
\frac{\bf q}{2}\bigg)  \phi^\star\bigg({\bf P} +\frac{\bf q}{2}\bigg)  e^{ - i {\bf q}\cdot {\bf  r}} .
\label{rhoint2}
\end{eqnarray}

The matrix elements of the EMT in the localized states with $R\to 0$ can be obtained analogously
to the electromagnetic case. As we will see below, the leading order contributions to
$t^{00}_{\phi}({\bf r}) $ and $t^{0i}_{\phi}({\bf r}) $ 
are of the order of $1/R$, and the $t^{ij}_{\phi}({\bf r}) $ terms need to be treated differently
from the others, when expanding in $R$. The reason for that is that, the components of $t^{ij}_{\phi}({\bf r})$, unlike $t^{00}_{\phi}({\bf r})$
and $t^{0i}_{\phi}({\bf r})$, which contain only information about the energy and spin densities,
respectively, encode information about the internal
pressure and shear forces as well as about the motion of the system
\cite{Freese:2021mzg,Panteleeva:2022uii}. That is, $t^{i j}_{\phi}({\bf r})$ needs to be decomposed
to a component $t^{ij}_{\phi,0}({\bf r})$  that describes the motion of the system as whole,
and a component that encodes information about pressure and shear forces $t^{ij}_{\phi,2}({\bf r})$.
Therefore, after expanding in $R$, we keep the leading order contribution of each of these terms.
The resulting expressions have the form:
\begin{subequations}
\label{tuv}
\begin{eqnarray}
t^{00}_{\phi}(s',s,{\bf r})   
 &=& N_{\phi,R} \int \frac{d^3 {q}}{(2\pi)^3 }\, e^{ - i {\bf q}\cdot {\bf  r}} \int d^2 \hat{n}  \Biggl\{   \mathcal{E}_0(q_{\perp}^{2})~\delta_{s' s} + \left[ \mathcal{E}_1(q_{\perp}^{2}) ~ \hat n^k\hat n^l  
  +  \mathcal{E}_2(q_{\perp}^{2})~  \frac{q^k_\perp q^l_\perp}{{m^2}}  \right]  {\hat Q^{k l}_{s' s}} \Biggl\} , \\
 \label{t00}
 t^{0i}_{\phi}(s',s,{\bf r})  
&  = & i \, N_{\phi,R} \int \frac{ d^3 {q}}{(2\pi)^3 } \, e^{ - i {\bf q}\cdot {\bf  r}} \int d^2 \hat{n}  \Biggl\{ ~\left[ \mathcal{C}_0(q_{\perp}^{2})  \, \epsilon^{k l n} \hat n^l \hat n^i    +  \mathcal{C}_1(q_{\perp}^{2})\, \epsilon^{i l n} \left( \delta^{kl} - \hat n^k  \hat n^l \right) \right] \, \frac{q^n_\perp}{m} \, {\hat S^{k}_{s' s}} 
      \nn 
&+& \Biggl[ \left(  \mathcal{C}_2(q_{\perp}^{2}) \hat n^t \hat n^z +  \mathcal{C}_3(q_{\perp}^{2}) \frac{q_\perp^t q_\perp^z}{{m^2}}  \right)~ \epsilon^{k l n}   \hat n^l \hat n^i  
\nn
& +& \left(\mathcal{C}_4(q_{\perp}^{2})\hat n^t \hat n^z  +\mathcal{C}_5(q_{\perp}^{2})s\frac{q_\perp^t q_\perp^z}{{m^2}} \right) \epsilon^{i l n} \left( \delta^{kl} - \hat n^k \hat n^l \right)  \Biggl] \frac{q^n_\perp}{m} \,{\hat O^{k t z}_{s' s}} \Biggl\} ,  \\
\label{t0i}
t^{ij}_{\phi}(s',s,{\bf r})  &=&t^{ij}_{\phi,0}(s',s,{\bf r}) +   t^{ij}_{\phi,2}(s',s,{\bf r}),
\end{eqnarray}
\end{subequations} 
where
\begin{subequations}
\begin{eqnarray}
 t^{ij}_{\phi,0}(s',s,{\bf r})  &=&  N_{\phi,R}  \int \frac{ d^3 {q}}{(2\pi)^3 }~e^{ - i {\bf q}\cdot {\bf  r}} \int d^2 \hat{n} \,  \hat n^i   \hat n^j \Biggl\{   \mathcal{E}_0(q_{\perp}^{2})~\delta_{s' s} 
 + \left[ \mathcal{E}_1 (q_{\perp}^{2}) ~ \hat n^k\hat n^l   +  \mathcal{E}_2(q_{\perp}^{2})~  \frac{q^k_\perp q^l_\perp}{{m^2}}  \right]  {\hat Q^{k l}_{s' s}} \Biggl\} \,  \\ 
 t^{ij}_{\phi,2}(s',s,{\bf r}) &=& N_{\phi,R,2}   \int \frac{ d^3 {q}}{(2\pi)^3 }  \, e^{ - i {\bf q}\cdot {\bf  r}} \int d^2 \hat{n} 
 \nn
 &\times& \Biggl\{ \frac{1}{2 m^2 }
\left(  q^i q^j- { q}_\perp^2 \delta^{i j } \right)   \Biggl[\mathcal{W}_0(q_{\perp}^{2}) \delta_{s' s} +\Biggl[\mathcal{W}_1(q_{\perp}^{2}) \hat n^k \hat n^l  
+\mathcal{W}_2(q_{\perp}^{2}) \frac{q_\perp^k q_\perp^l}{{m^2}} \Biggl]  {\hat Q^{kl}_{s' s}} \Biggl] \nn 
&-&
2\delta^{i j } \Biggl[ \mathcal{U}_0 (q_{\perp}^{2}) \delta_{s' s} + \left[ \mathcal{U}_1(q_{\perp}^{2}) \hat n^k \hat n^l + \mathcal{U}_2 (q_{\perp}^{2}) \frac{q_\perp^k q_\perp^l}{{m^2}} \right]  {\hat Q^{k l }_{s' s}}  \Biggl] \Biggl\}  \ ,
\label{rhoint2NRGR}
\end{eqnarray}
\end{subequations} 
and the coefficient functions $\mathcal{E}_i,  \mathcal{C}_i, \mathcal{W}_i$ and $\mathcal{U}_i$
are given in the appendix and further,
\bea
N_{\phi,R} &=& \frac{1}{R}  \int_0^\infty  \, d{ Q}\, { Q}^3 |\tilde\phi(|{\bf Q}|)|^2 
  \,,\nonumber\\
  N_{\phi,R,2} &=& \frac{m^2 R}{2}  \int_0^\infty  \, d Q\, Q |\tilde\phi(|{\bf Q}|)|^2
  \,.
  \label{normaliz}
\eea

Below, we specify the multipole expansion of $t^{\mu\nu}_{\phi}(s',s,{\bf r})$. For the $00$th
component the result reads
\begin{eqnarray}
t^{00}_{\phi}(s',s,{\bf r})
 &=&N_{\phi,R} \Biggl\{ \rho^E_0  \left(r\right)\delta_{s's}+\rho^E_2 \left(r\right)Y_{2}^{kl}\left(\Omega_{r}\right)\hat{Q}_{s's}^{kl}\Biggl\},
 \label{t00_Multipole}
\end{eqnarray}
where the monopole and quadrupole energy densities are identified as $\rho^E_0$ and $\rho^E_2$,
respectively. Their expressions have the form:
\begin{eqnarray}
\rho^E_{0} \left(r\right)&=& \int\frac{d^{3}q}{(2\pi)^{3}}\,e^{-i{\bf q}\cdot{\bf r}}\int d^{2}\hat{n}\bigg\{ F_{1,0}(-q_\perp^2) - \frac{2}{3} F_{6,0}(-q_\perp^2) 
\nn
&+& \frac{q_\perp^2}{3m^2} \left[ -F_{1,0}(-q_\perp^2) + \frac{1}{2} F_{1,1}(-q_\perp^2) + F_{4,0}(-q_\perp^2) +2 F_{5,0}(-q_\perp^2) \right] 
 -  \frac{q_\perp^4}{12m^4} \left[ F_{1,1}(-q_\perp^2) - F_{4,1}(-q_\perp^2) \right]  \bigg\},
\\
\rho^E_{2} \left(r\right)&=&-r\frac{d}{dr}\frac{1}{r}\frac{d}{dr}\int\frac{d^{3}q}{(2\pi)^{3}}\,e^{-i{\bf q}\cdot{\bf r}}\int d^{2}\hat{n} \frac{1}{q^{2}} \bigg\{ \frac{2}{3}F_{6,0}(-q_\perp^2)  -\frac{q_\perp^2}{q^2} F_{6,0}(-q_\perp^2)
\nn
&+&  \frac{q_\perp^2}{12 m^2} \left[ 4 F_{1,0}(-q_\perp^2)  +F_{1,1}(-q_\perp^2) - 4 F_{4,0}(-q_\perp^2) - 8 F_{5,0}(-q_\perp^2)   \right] + \frac{q_\perp^4}{12 m^4}  \left[ F_{1,1}(-q_\perp^2) - F_{4,1}(-q_\perp^2)  \right]
\nn
&+&  \frac{q_\perp^4}{4 m^2q^2} \left[4F_{5,0}(-q_\perp^2) -F_{1,0}(-q_\perp^2) -F_{1,1}(-q_\perp^2)  \right]   - \frac{q_\perp^6}{16 m^4 q^2}F_{1,1}(-q_\perp^2) \ \bigg\} \,.
\label{t00_Etilde}
\end{eqnarray}
For the $0i$th components we have
\begin{eqnarray}
 t^{0i}_{\phi}(s',s,{\bf r})  
&=& N_{\phi,R} \biggl[ ~\epsilon^{ikn}\hat{S}_{s's}^{k}Y_{1}^{n} \frac1r \rho^J_1 \left(r\right) \text{ }+ ~\epsilon^{ikn}\hat{O}_{s's}^{ktz}Y_{3}^{ntz}\, \frac1r \rho^J_3 \left(r\right) \biggl],
\label{t0i_Multipole}
\end{eqnarray}
where
\begin{eqnarray}
\rho^J_1 \left(r\right)&=& \frac rm \frac{d}{dr} \int\frac{d^{3}q}{(2\pi)^{3}}\,e^{-i{\bf q}\cdot{\bf r}}\int d^{2}\hat{n} \frac{q_{\perp}^{2}}{2q^{2}} \bigg\{ F_{1,0}(-q_\perp^2) -\frac{2}{3}F_{4,0}(-q_\perp^2)-\frac{4}{15}F_{6,0}(-q_\perp^2) 
\nonumber \\
&+&  \frac{q_\perp^2}{15 m^2} \left[  - F_{1,0}(-q_\perp^2) +\frac{7}{2} F_{1,1}(-q_\perp^2)+  2 F_{4,0}(-q_\perp^2) - 2 F_{4,1}(-q_\perp^2)  + 4 F_{5,0}(-q_\perp^2) \right]    \nn 
&+&
 \frac{q_\perp^4}{60 m^4} \left[ - F_{1,1}(-q_\perp^2) +  F_{4,0}(-q_\perp^2) +  F_{4,1}(-q_\perp^2) \right] \bigg\},
\\
\rho^J_3 \left(r\right)&=& \frac{r^{4}}{m}\left(\frac{1}{r}\frac{d}{dr}\right)^{3} \int\frac{d^{3}q}{(2\pi)^{3}}e^{-i{\bf q}\cdot{\bf r}}\int d^{2}\hat{n}\frac{q_{\perp}^{2}}{12q^{4}}\bigg\{-4F_{6,0}^{a}(-q_{\perp}^{2})+5\frac{q_{\perp}^{2}}{q^{2}}F_{6,0}^{a}(-q_{\perp}^{2})
\nn 
&-&\frac{q_{\perp}^{2}}{4m^{2}}\left[4F_{1,0}^{a}(-q_{\perp}^{2})+F_{1,1}^{a}(-q_{\perp}^{2})-8F_{4,0}^{a}(-q_{\perp}^{2})-2F_{4,1}^{a}(-q_{\perp}^{2})-16F_{5,0}^{a}(-q_{\perp}^{2})\right]
\nn
&+&\frac{5q_{\perp}^{4}}{4m^{2}q^{2}}\left[F_{1,0}^{a}(-q_{\perp}^{2})+F_{1,1}^{a}(-q_{\perp}^{2})-2F_{4,0}^{a}(-q_{\perp}^{2})-2F_{4,1}^{a}(-q_{\perp}^{2})-4F_{5,0}^{a}(-q_{\perp}^{2})\right]
\nn
&-&\frac{q_{\perp}^{4}}{4m^{4}}\left[F_{1,1}^{a}(-q_{\perp}^{2})-2F_{4,1}^{a}(-q_{\perp}^{2})\right]+\frac{5q_{\perp}^{6}}{16m^{4}q^{2}}\left[F_{1,1}^{a}(-q_{\perp}^{2})-2F_{4,1}^{a}(-q_{\perp}^{2})\right]\bigg\}.
\end{eqnarray}

The spin density is given by
\begin{eqnarray}
J^{i}(\bm{r},s',s)&=& \epsilon^{i j k } r^j  t^{0k}_{\phi}(s',s,{\bf r}) =N_{\phi,R}\,\Biggl\{\left(\frac{2}{3}\delta^{il}Y_{0}-Y_{2}^{il}\right)\rho^J_1\left(r\right)\text{ }\hat{S}_{s's}^{l}\nn
&&+\left[-Y_{4}^{iltz}+\frac{2}{35}\left(8\delta^{il}Y_{2}^{tz}+\delta^{it}Y_{2}^{lz}+\delta^{iz}Y_{2}^{lt}\right)\right] 
\rho^J_3\left(r\right)\,\hat{O}_{s's}^{ltz}\Biggl\}.
\end{eqnarray}
We identify the monopole angular momentum density $(J_{\text{mono}}^i)$ as the term with the
structure $Y_{0}\hat{S}_{s's}^{i} $, i.e. 
\begin{eqnarray}
J_{\text{mono}}^i\left({\bf r}, s', s\right) &=& \frac{2}{3}  N_{\phi,R}\;  \hat{S}_{s's}^{i}   \rho^J_1\left(r\right) ,
\end{eqnarray}
and, following Ref.~\cite{Kim:2020lrs}, the averaged angular momentum density is given by
\begin{eqnarray}
\rho_{J}\left(r\right)&\equiv&\frac{S}{\text{Tr}\left[\hat{S}^{2}\right]}\sum_{s^{\prime},s,i}S_{ss^{\prime}}^{i}J_{\text{mono}}^{i}\left(\vec{r},s^{\prime},s\right)=N_{\phi,R} \, \rho^J_1\left(r\right) \ ,
\end{eqnarray}
 where $S=3/2$ is the spin of the system.

\medskip

Finally, for the $ij$th components we obtain
\begin{eqnarray}
 t^{ij}_{\phi,0}(s',s,{\bf r})  &=& N_{\phi,R}\bigg\{\left[ {a}_{1}\left(r\right)\delta^{ij}-\left(\frac{\delta^{ij}}{3}\partial^{2}+Y_{2}^{ij}r\frac{d}{dr}\frac{1}{r}\frac{d}{dr}\right) 
 {a}_{2}\left(r\right)\right]\delta_{s's} +\hat{Q}_{s's}^{ij} {a}_{3} \left(r\right)
\nn
&&+\hat{Q}_{s's}^{kl}\partial_{k}\partial_{l}\left(\frac{\delta^{ij}}{3}\partial^{2}+Y_{2}^{ij}r\frac{d}{dr}\frac{1}{r}\frac{d}{dr}\right){a}_{4}\left(r\right)
-\delta^{ij}\hat{Q}_{s's}^{kl}\partial_{k}\partial_{l}{a}_{5}\left(r\right)
\nn
&&-\left[\frac{2}{3}\hat{Q}_{s's}^{ij}\partial^{2}+\left(\hat{Q}_{s's}^{iv}Y_{2}^{jv}+\hat{Q}_{s's}^{jv}Y_{2}^{iv}\right)r\frac{d}{dr}\frac{1}{r}\frac{d}{dr}\right]{a}_{6}\left(r\right) \bigg\},\label{tij_multipole} \\
 t^{ij}_{\phi,2}(s',s,{\bf r}) 
&=&
N_{\phi,R,2}\bigg\{
-\frac{1}{2m^{2}}\left(\frac{\delta^{ij}}{3}\partial^{2}+Y_{2}^{ij}r\frac{d}{dr}\frac{1}{r}\frac{d}{dr}\right){w}_{0} \left(r\right) \, \delta_{s's}\nn
&&+\frac{1}{2m^{2}}\hat{Q}_{s's}^{kl}\partial_{k}\partial_{l}\left[\left(\frac{\delta^{ij}}{3}\partial^{2}+Y_{2}^{ij}r\frac{d}{dr}\frac{1}{r}\frac{d}{dr}\right)
 {w}_{1}\left(r\right) 
\right]\nn
&&+\delta^{ij}\left[{v}_{0}\left(r\right)~\delta_{s's}-\hat{Q}_{s's}^{kl}\partial_{k}\partial_{l}
\,  {v}_{1}\left(r\right) 
\right]\bigg\},
\label{tij_multipole_F2F3}
\end{eqnarray}
where the coefficient functions ${a}(r)$, $w(r)$ and $v(r)$ are given in the appendix.

To obtain the pressure and shear force densities we consider a conserved EMT    
and take the  part of  ${\tilde t}^{ij}_{\phi,2}(s',s,{\bf r})$ linear in $R$ (where the
tilde means only conserved EMTs are considered),
which we parametrize as follows \cite{Polyakov:2019lbq}:
\begin{eqnarray}
{\tilde t}^{ij}_{\phi,2}(s',s,{\bf r}) &=& N_{\phi,R,2} \Biggl\{ p_0(r) \delta^{i j } \delta_{s' s} + s_0(r) Y_{2}^{ij} \delta_{s' s} + p_2(r) \hat{Q}_{s's}^{i j} + 2 s_{2}(r) 
\left[  \hat{Q}_{s's}^{i k} Y_{2}^{k j} +  \hat{Q}_{s's}^{j k} Y_{2}^{k i} - \delta^{i j } \hat{Q}_{s's}^{k l } Y_{2}^{k l} \right]   \nn 
&-& \frac{1}{m^2}  \hat{Q}_{s's}^{k l}  \partial_k \partial_l \left[ p_3(r) \delta^{i j } + s_3(r) Y_{2}^{i j}  \right] \Biggl\} , \label{ijp}
\end{eqnarray}
where  
$p_0(r)$ and $s_0(r)$ are the pressure and shear force densities also appearing in the
spherically symmetric hadrons, respectively, $p_2(r)$, $p_3(r)$ correspond to the quadrupole
pressure densities, 
and $s_2(r)$, $s_3(r)$ are the quadrupole shear force densities.\footnote{Another equivalent parametrization is given in Ref.~\cite{Panteleeva:2020ejw}, where the normal and tangential forces can be defined in a compact way. However, it has been shown that the parametrization of Ref.~\cite{Polyakov:2019lbq} has advantages in studying the mechanical structure, whenever performing an Abel transformation is involved \cite{Kim:2022wkc}.} 
Comparing Eqs.~(\ref{ijp}) and (\ref{tij_multipole_F2F3}) we obtain for the pressure and shear
forces the following results: 
\begin{eqnarray}
p_{0}(r)&=&\tilde{v}_{0}\left(r\right)-\frac{1}{6m^{2}}\partial^{2}{w}_{0}\left(r\right),\quad s_{0}(r)=-\frac{1}{2m^{2}}r\frac{d}{dr}\frac{1}{r}\frac{d}{dr}{w}_{0}\left(r\right) , \nonumber
\\
p_{2}(r)&=&0,\qquad \qquad  \qquad  \qquad  \qquad  s_{2}(r)=0 \label{s2p20} , \nonumber
\\
p_{3}(r)&=&m^{2}\tilde{v}_{1}\left(r\right)-\frac{1}{6}\partial^{2}{w}_{1}\left(r\right),\quad s_{3}(r)=-\frac{1}{2}r\frac{d}{dr}\frac{1}{r}\frac{d}{dr}{w}_{1}\left(r\right) ,
\label{defpands}
\end{eqnarray}
where  the coefficient functions ${ w}(r)$ and ${\tilde v}(r)$ are  given as
\begin{eqnarray}
{w}_{0}\left(r\right)&=&\int \frac{d^{2}\hat{n} d^{3}q}{\left(2\pi\right)^{3}}e^{-i\vec{q}\cdot\vec{r}} \Biggl[ F_{2,0}(-q_\perp^2) + \frac{q_\perp^2}{6m^2} \left[ -2 F_{2,0}(-q_\perp^2) + F_{2,1}(-q_\perp^2) \right] - \frac{q_\perp^4}{12m^4} F_{2,1}(-q_\perp^2) \Biggl] , \nonumber \\
\tilde{v}_{0}\left(r\right)&=&\int\frac{d^{2}\hat{n}d^{3}q}{\left(2\pi\right)^{3}}e^{-i\vec{q}\cdot\vec{r}}\left(-\frac{q_{\perp}^{2}}{2m^{2}}\right)\left[F_{2,0}(-q_{\perp}^{2})+\frac{q_{\perp}^{2}}{6m^{2}}\left[-2F_{2,0}(-q_{\perp}^{2})+F_{2,1}(-q_{\perp}^{2})\right]-\frac{q_{\perp}^{4}}{12m^{4}}F_{2,1}(-q_{\perp}^{2})\right], \nonumber 
\\
{w}_{1}\left(r\right)
&=&\int\frac{d^{2}\hat{n}d^{3}q}{\left(2\pi\right)^{3}}e^{-i\vec{q}\cdot\vec{r}}\frac{q_{\perp}^{2}}{2m^{2}q^{2}}\left[\left(\frac{2}{3}-\frac{q_{\perp}^{2}}{2q^{2}}\right)F_{2,0}(-q_{\perp}^{2})+\frac{1}{2}\left(\frac{1}{3}+\frac{q_{\perp}^{2}}{3m^{2}}-\frac{q_{\perp}^{2}}{q^{2}}-\frac{q_{\perp}^{4}}{4m^{2}q^{2}}\right)F_{2,1} (-q_{\perp}^{2})\right]  , \nonumber 
\\
\tilde{v}_{1}\left(r\right)&=&\int\frac{d^{2}\hat{n}d^{3}q}{\left(2\pi\right)^{3}}e^{-i\vec{q}\cdot\vec{r}}\left(-\frac{q_{\perp}^{4}}{4m^{4}q^{2}}\right)\left[\left(\frac{2}{3}-\frac{q_{\perp}^{2}}{2q^{2}}\right)F_{2,0}(-q_{\perp}^{2})+\frac{1}{2}\left(\frac{1}{3}+\frac{q_{\perp}^{2}}{3m^{2}}-\frac{q_{\perp}^{2}}{q^{2}}-\frac{q_{\perp}^{4}}{4m^{2}q^{2}}\right)F_{2,1}(-q_{\perp}^{2})\right].
\label{defwandv}
\end{eqnarray}
It is clear from Eqs.~(\ref{defpands}) and (\ref{defwandv}) that
the pressure and shear forces are expressed in terms of $F_{2,0}$ and $F_{2,1}$ only.\footnote{Notice
that the feature $p_2=s_2=0$ is not identical with the result of the large $N_c$ limit for baryons
in the chiral soliton model as obtained in Ref.~\cite{Panteleeva:2020ejw}, where another
parameterization of the pressure and shear forces
is used. This can be easily checked by converting the pressure and shear forces in
Eq.~(\ref{defpands}) to the notations used in Ref.~\cite{Panteleeva:2020ejw}.}

\medskip

Below we obtain the differential equation for the pressure and shear forces that follows from
the conservation of the EMT. In that case,  we have 
to make our matrix element in Eq.~(\ref{EMTdefJH})  time-dependent, i.e. we substitute
$e^{ - i {\bf q}\cdot {\bf  r} } \mapsto  e^{ i q_0 t- i {\bf q}\cdot {\bf  r} }$.~
The conservation of the  EMT leads to
\begin{eqnarray} \label{conservation_of_EMT}
\partial_\mu t^{\mu\nu}_{\phi}(s',s,{\bf r},t) |_{t=0} =   \partial_0 t^{0\nu}_{\phi}(s',s,{\bf r},t) |_{t=0} +  \partial_i t^{i\nu}_{\phi}(s',s,{\bf r},t) |_{t=0} =0.
\end{eqnarray}
Notice that $\partial_0$ and $ |_{t=0}$ do not commute, i.e. to obtain $ \partial_0
t^{0\nu}_{\phi}(s',s,{\bf r},t) |_{t=0}$ we first take the derivative of $t^{0\nu}_{\phi}(s',s,{r})$,
then put $t=0$ and after that perform the expansion around  $R= 0$. Since we are
interested in the pressure and shear forces, we consider the case with $\nu=j$ and keep
all contributions linear  in $R$. We obtain the following differential equation,
adhering to the notation of Ref.~\cite{Polyakov:2019lbq},
\begin{eqnarray}
\label{consEq}
p_{n}^{\prime}(r)+\frac{2}{3}s_{n}^{\prime}(r)+\frac{2}{r}s_{n}(r) = {h}_{n}^{\prime}\left(r\right) , \ \ \ \text{with} \; n=0,2,3 , 
\end{eqnarray} 
where
\begin{eqnarray}
{h}_{0}\left(r\right)&=&\int\frac{d^{3}q}{\left(2\pi\right)^{3}}e^{-i{\bf q}\cdot{\bf r}} \int d^{2}\hat{n} \frac{\left({\bf q}\cdot\hat{n}\right)^{2}}{2m^{2}}\mathcal{W}_{0}\left(q_{\perp}^{2}\right) , \quad {h}_{2}\left(r\right) =0,
\nonumber
\\
{h}_{3}\left(r\right)&=&\int\frac{d^{3}q}{\left(2\pi\right)^{3}}e^{-i{\bf q}\cdot{\bf r}} \int d^{2}\hat{n} \frac{\left({\bf q}\cdot\hat{n}\right)^{2}}{2q^{2}}\left[\mathcal{W}_{1}\left(q_{\perp}^{2}\right)\;\left(1-\frac{3q_{\perp}^{2}}{2q^{2}}\right)+\mathcal{W}_{2}\left(q_{\perp}^{2}\right)~\frac{q_{\perp}^{2}}{2m^{2}}\left(\frac{3q_{\perp}^{2}}{q^{2}}-1\right)\right] ,
\end{eqnarray}
with $r=|{\bf r}|$, and the coefficient functions $\mathcal{W}_{0,1,2}(q^2_\perp)$ are given in the
Appendix. In contrast, there are no $h_n(r)$ terms in Eq.~(\ref{consEq}) in the case of the
Breit-frame because of the absence of the temporal dependence in $t^{0\nu}$ component (due to $q^0=0$).

The pressure densities $p_n(r)$ comply with the von Laue stability condition
\begin{eqnarray}
\int d^{3}r\,p_{n}(r)&=&0,  \ \ \   \mathrm{with}\ \ n=0,2,3, 
\end{eqnarray} 
as long as $\text{lim}_{q^2_\perp\rightarrow0} \left(q_{\perp}^{2}\right)^{\delta}F_{2,0}\left( -q^2_\perp \right)=0$ and $\text{lim}_{q^2_\perp\rightarrow0} \left(q_{\perp}^{2}\right)^{\delta}F_{2,1}\left( -q^2_\perp \right)=0$, for $\delta>0$.

The dimensionless constants (generalized D-terms) are defined by
\begin{eqnarray}
\mathcal{D}_{n}
	&=& {- \frac{4}{15}\,m^2 \int{\rm d}^3r\,r^2s_n(r)} 
	= m^2\int d^{3}r\,r^{2}\left[p_{n}\left(r\right)-{h}_{n}\left(r\right)\right],  \ \ \   \mathrm{with}\ \ n=0,2,3.
\end{eqnarray} 
Note that the above definition differs from that of the Breit-frame
case~\cite{Panteleeva:2020ejw} by the ${h}_{n}\left(r\right)$ terms. 

\medskip

The spherical components of the internal forces ($dF_{r},dF_{\theta}$ and $dF_{\varphi}$)  acting on the radial area element ($d\bm{S}=dS_{r}\hat{\bm{e}}_{r}+dS_{\theta}\hat{\bm{e}}_{\theta}+dS_{\varphi}\hat{\bm{e}}_{\varphi}$) are expressed as follows
\begin{eqnarray}
\frac{dF_{r}}{dS_{r}}&=&N_{\phi,R,2}\Bigg[\left(p_{0}(r)+\frac{2}{3}s_{0}(r)\right)\delta_{s^{\prime}s}+\left(p_{2}(r)+\frac{2}{3}s_{2}(r)\right)\hat{Q}_{s^{\prime}s}^{rr}\\&&-\frac{1}{m^{2}}\hat{Q}_{s^{\prime}s}^{rr}\left(r\frac{d}{dr}\frac{1}{r}\frac{d}{dr}\left(p_{3}(r)+\frac{2}{3}s_{3}(r)\right)+s_{3}(r)\frac{2}{r^{2}}\right)\Bigg] ,
\\
\frac{dF_{\theta}}{dS_{r}}&=&N_{\phi,R,2}\bigg[\left(p_{2}(r)+\frac{2}{3}s_{2}(r)\right)\hat{Q}_{s^{\prime}s}^{\theta r}-\frac{2}{m^{2}}\hat{Q}_{s^{\prime}s}^{\theta r}\frac{d}{dr}\frac{s_{3}(r)}{r}\bigg] ,
\\
\frac{dF_{\varphi}}{dS_{r}}&=&N_{\phi,R,2}\bigg[\left(p_{2}(r)+\frac{2}{3}s_{2}(r)\right)\hat{Q}_{s^{\prime}s}^{\varphi r}-\frac{2}{m^{2}}\hat{Q}_{s^{\prime}s}^{\varphi r}\frac{d}{dr}\frac{s_{3}(r)}{r}\bigg] .
\end{eqnarray} 
Notice that for an unpolarized spin-3/2 hadron, the normal force acting on the radial
area element (${dF_{r}}/{dS_{r}}$) is solely due to $p_{0}(r)+\frac{2}{3}s_{0}(r)$.

As defined in Ref.~\cite{Polyakov:2018zvc}, the mechanical radius is given by
\begin{eqnarray}
\langle r_{n}^{2} \rangle_{\mathrm{mech}} = \frac{\int d^{3}r \, r^{2} \left[p_{n}(r) + \frac{2}{3}s_{n}(r)\right]}{\int d^{3}r \, \left[p_{n}(r) + \frac{2}{3}s_{n}(r)\right]}\,.
\label{eq:mech}
\end{eqnarray}
Notice that Eqs.~(\ref{defpands}),  (\ref{defwandv}) and (\ref{eq:mech}) lead to
expressions for the radii which differ from those of the Breit frame.

Analogously to the case of the electromagnetic current, the static approximation is obtained by
first expanding the integrand in $1/m$ and after that taking the limit of a sharply localized
wave packet.  The resulting naive densities read:
\begin{eqnarray}
t^{00}_{\text{naive}}(s',s,{\bf r})  &=& \int \frac{d^3q}{(2\pi)^3} e^{ - i {\bf q}\cdot {\bf  r}} \Biggl\{ \left[ m  F_{1,0} \left( - {\bf q}^2 \right) +  \frac{ {\bf q}^2 }{6 m}F_{1,1} \left( - {\bf q}^2 \right)\right] \delta_{s' s}- \frac{1}{6 m } F_{1,1} \left( - {\bf q}^2 \right)  q^k q^n   {\hat Q^{ k n }_{s' s}}  \Biggl\}\,,\nn
t^{0i}_{\text{naive}}(s',s,{\bf r})  &=&   -i \int \frac{d^3q}{(2\pi)^3} e^{ - i {\bf q}\cdot {\bf  r}} \epsilon^{i l k  } q^l  \Biggl\{  \frac{1 }{3}\left[  F_{4,0}\left(- {\bf q}^2 \right) +\frac{{\bf q}^2}{5 m ^2} F_{4,1} \left(- {\bf q}^2 \right) \right] {\hat S^{ k  }_{s' s}} - \frac{  F_{4,1} \left(- {\bf q}^2 \right) }{6 m ^2}  q^n q^t {\hat O^{ k n t}_{s' s}} \Biggl\} ,
\nn
t^{ij}_{\text{naive}}(s',s,{\bf r})  &=& \int \frac{d^3 n}{ R^2} |\tilde\phi({|
  {\bf n}|})|^2 n^i n^j \int \frac{d^3q}{(2\pi)^3} e^{ - i {\bf q}\cdot {\bf  r}} \bigg\{  \left[ \frac{1}{m}  F_{1,0} \left( - {\bf q}^2 \right) +  \frac{ {\bf q}^2 }{6 m^3}F_{1,1} \left( - {\bf q}^2 \right)\right] \delta_{s' s} 
  \nn &&- \frac{1}{6 m^3 } F_{1,1} \left( - {\bf q}^2 \right)  q^k q^n   {\hat Q^{ k n }_{s' s}}   \bigg\} 
 \nn
 &+& \int \frac{d^3q}{(2\pi)^3}  e^{ - i {\bf q}\cdot {\bf  r}}  \bigg\{ \left(q^i q^j -{\bf q}^2 \delta^{i j}\right)\bigg[ \left( \frac{1}{4m}  F_{2,0} \left( - {\bf q}^2 \right) 
 +  \frac{ {\bf q}^2 }{24 m^3}F_{2,1} \left( - {\bf q}^2 \right)\right) \delta_{s' s} 
 \nn
 &&- \frac{1}{24 m^3 } F_{2,1} \left( - {\bf q}^2 \right)  q^k q^n   {\hat Q^{ k n }_{s' s}} \bigg] - \frac{4}{ 3 m}F_{5,0} \bigg[  \left(q^i q^j -{\bf q}^2 \delta^{i j}\right)\delta_{s' s} 
 \nn 
 &&+ \frac{1}{2} \left( q^2 {\hat Q^{ i j }_{s' s}} + \delta^{i j} q^k q^n {\hat Q^{ k n }_{s' s}}  -q^k q^i {\hat Q^{ k j }_{s' s}}-q^k q^j {\hat Q^{ k i }_{s' s}}  \right) \bigg] 
 + \frac{F_{3,1} \left( - {\bf q}^2 \right) }{6 m } \delta^{i j }q^k q^n  {\hat Q^{ k n }_{s' s}}
\nn 
&&- \left(  m F_{3,0} \left( - {\bf q}^2 \right) + \frac{{\bf q}^2}{6 m } F_{3,1} \left( - {\bf q}^2 \right) +\frac{2 m}{3 } F_{6,0} \left( - {\bf q}^2 \right) \right) \delta^{i j }\delta_{s' s} 
+ \frac{ 2 m  }{3} F_{6,0} \left( - {\bf q}^2 \right)  {\hat Q^{ i j }_{s' s}} 
\bigg\} .
\label{tnaive}
  \end{eqnarray} 
The above results for $t^{00}_{\text{naive}}(s',s,{\bf r})$ and $t^{0i}_{\text{naive}}(s',s,{\bf r})$ agree
with the corresponding Breit-frame expressions of Ref.~\cite{Kim:2020lrs} modulo terms of
higher-orders in the $1/m$ expansion, contained in the Breit-frame expressions. On the other hand,
for the $ij$th component, Ref.~\cite{Kim:2020lrs} only quotes the result corresponding to
the second term of $t^{ij}_{\text{naive}}(s',s,{\bf r})$, contained in the last four lines of
Eq.~(\ref{tnaive}), which is to be interpreted as characterizing the internal forces of
the considered system \cite{Freese:2021mzg,Panteleeva:2022uii}. Again, the two expressions
agree modulo terms of higher-orders in the $1/m$ expansion, contained in the Breit-frame
expression of Ref.~\cite{Kim:2020lrs}.

\section{Large distance behavior of the energy, spin, pressure and shear forces distributions of the delta resonance in chiral EFT }
\label{LDB} 
 
In Ref.~\cite{Alharazin:2022wjj} the GFFs of the delta resonance have been calculated up to
third chiral order. Using these results and restricting ourselves to the 
region of distances $1/\Lambda_{\text{strong}} \ll r \ll 1/M_\pi$ 
we obtained approximate asymptotic behavior of the spatial densities 
from the singularities of GFFs at $t=0$ in the chiral limit. The third 
order one-loop calculations of Ref. \cite{Alharazin:2022wjj} led to the following
expressions for corresponding leading non-analytical  parts of the GFFs: 
\begin{eqnarray}
&& F_{\rm 1,0}(t) = -\frac{25 g_1^2 }{9216 F^2 m_\Delta}~t\,\sqrt{-t} 
  \, , \nonumber
\\
&& F_{\rm 1,1}(t) =-\frac{5 g_1^2 m_\Delta}{1536 F^2 }\,\sqrt{-t} 
   \, ,  \nonumber
\\
&& F_{\rm 2,0}(t) = \frac{5 g_1^2 m_\Delta}{768 F^2}\,\sqrt{-t} 
    \,, \nonumber
\\
&&F_{\rm 2,1}(t)= \frac{5 g_1^2 m_\Delta^3}{384 F^2}\,\frac{\sqrt{-t}}{t}\, ,\nonumber\\ 
&& F_{\rm 4,0}(t) = -\frac{5 g_1^2}{1728 \pi ^2 F^2}~t \ln(-t/ m_N^2) 
\,, \nonumber
\end{eqnarray}
\begin{eqnarray}
&& F_{\rm 4,1}(t) =0\, ,  \nonumber\\
&& F_{\rm 5,0}(t) =  -\frac{5 g_1^2 m_\Delta}{9216 F^2}\,\sqrt{-t} \, ,
\label{non-a}
\end{eqnarray}
where $t=q^2$. Using the above results we obtain the following long-range behavior for
the densities derived in the previous section
 \begin{eqnarray}
\rho^E_0(r) &=& \frac{25 g_1^2}{1536 F^2 m_\Delta}\frac{1}{r^6}-\frac{10 g_1^2}{81 \pi ^2 F^2 m_\Delta^2}\frac{1}{r^7}  + {\cal O}\left(\frac{1}{r^8} \right)
\,,\label{energy0}\\
\rho^E_2(r) &=& \frac{35 g_1^2}{6144 F^2 m_\Delta}\frac{1}{r^6}+\frac{35 g_1^2}{162 \pi ^2 F^2 m_\Delta^2}\frac{1}{r^7} + {\cal O}\left(\frac{1}{r^8} \right)
\,,
\\
\rho^J_1(r) &=& \frac{5 g_1^2}{162 \pi^2 F^2 m_\Delta}\frac{1}{r^5} - \frac{125 g_1^2}{3072 F^2 m_\Delta^2}\frac{1}{r^6}  + {\cal O}\left(\frac{1}{r^7} \right)
\,, \\
\rho^J_3(r) &=& - \frac{625 g_1^2}{24576 F^2 m_\Delta^2}\frac{1}{r^6} + \frac{5 g_1^2}{54 \pi ^2 F^2 m_\Delta^3}\frac{1}{r^7}  + {\cal O}\left(\frac{1}{r^8} \right)
\,,\\
p_0(r) &=& -\frac{25 g_1^2}{2304 F^2 m_\Delta}\frac{1}{r^6}-\frac{75 g_1^2}{1024 F^2 m_\Delta^3}\frac{1}{r^8}\label{pre0} + {\cal O}\left(\frac{1}{r^9} \right) \,,\\
s_0(r) &=&  \frac{5 g_1^2}{96 F^2 m_\Delta}\frac{1}{r^6}+\frac{15 g_1^2}{64 F^2 m_\Delta^3}\frac{1}{r^8}\label{she0} + {\cal O}\left(\frac{1}{r^8} \right) \,,\\
p_3(r) &=& \frac{85 g_1^2 m_\Delta}{221184 F^2 }\frac{1}{r^4}-\frac{155 g_1^2}{196608 F^2 m_\Delta}\frac{1}{r^6} + {\cal O}\left(\frac{1}{r^8} \right) 
\,,\label{pre3}\\
s_3(r) &=& -  \frac{25 g_1^2 m_\Delta}{9216 F^2 }\frac{1}{r^4}+\frac{15 g_1^2}{4096 F^2 m_\Delta}\frac{1}{r^6} + {\cal O}\left(\frac{1}{r^8} \right)
\,.\label{she3}
\label{densities}
\end{eqnarray}
Notice that while the delta resonances are unstable particles, our expressions satisfy the
general stability conditions of Ref.~\cite{Polyakov:2018zvc}, i.e. $\rho^E_0(r)> 0 $ and
$\frac{2}{3} s_0(r) + p_0(r) > 0$. This result is in agreement with the observation of
Ref.~\cite{Polyakov:2018zvc} that the general stability conditions are necessary but not
sufficient for a system to be stable.

\section{Summary and conclusions}
\label{summary}

In this work we applied the novel definition of local spatial densities using
sharply localized wave packets \cite{Epelbaum:2022fjc} to spin-3/2 systems. 
Matrix elements of the electromagnetic current and the energy-momentum tensor in the ZAMF
were considered and 
integral representations of associated spatial densities in terms of form factors were derived.
Following Ref.~\cite{Jaffe:2020ebz}, the corresponding expressions in the Breit-frame were obtained
by first expanding the integrands in inverse powers of the mass of the system and then
taking the limit of sharply localized wave packets. This corresponds to considering  
packet sizes that are much larger than  the Compton wavelength of the system. To apply
the new definition as well as the Breit-frame formulas one needs to take the packet sizes
much smaller than any length scales characterizing internal structure of the system. This
makes clear that the Breit-frame spatial densities cannot be used for systems whose
Compton wavelengths and  the radii have comparable sizes \cite{Jaffe:2020ebz}.
However, the novel definition used here does not impose any lower bound on the size of the 
wave packet and therefore can be applied to any systems. 
 
Considering the spatial components of the matrix elements of the EMT we obtained the
expressions of the pressure and the shear forces inside the spin 3/2-systems. 
We also obtained a differential equation satisfied by these quantities due to the
conservation of the EMT. 

The formalism can be extended to the $\Delta\to N$ transition
form factors, however, this is not straightforward and requires
a separate investigation. Work along such lines is under way.

\acknowledgements

We thank J.~Panteleeva for numerous discussions and for checking some formulas.
This work was supported in part by BMBF (Grant No. 05P21PCFP1), by
DFG and NSFC through funds provided to the Sino-German CRC 110
“Symmetries and the Emergence of Structure in QCD” (NSFC Grant
No. 11621131001, DFG Project-ID 196253076 - TRR 110),
by CAS through a President’s International Fellowship Initiative (PIFI)
(Grant No. 2018DM0034), by the VolkswagenStiftung
(Grant No. 93562), by the EU Horizon 2020 research and
innovation program (STRONG-2020, grant agreement No. 824093),
and by the Heisenberg-Landau Program 2021, 
by Guangdong Provincial
funding with Grant
No. 2019QN01X172, the National Natural Science Foundation of China
with Grant No. 12035007 and No. 11947228, Guangdong Major Project of
Basic and Applied Basic Research No. 2020B0301030008, and the Department of Science and
Technology of Guangdong Province with Grant No. 2022A0505030010.

\appendix

\section{Coefficient functions} 
Coefficient functions for $j^{0}_{\phi}(s',s,{\bf r}) $ and $j^{i}_{\phi}(s',s,{\bf r}) $:
\begin{eqnarray}
\mathcal{Z}_0(-q_\perp^2)&=& F_{1,0}^{V}(-q_\perp^2)  + \frac{q_\perp^2}{6 m^2} \left[ -2 F_{1,0}^{V}(-q_\perp^2) +  F_{1,1}^{V}(-q_\perp^2) + F_{2,0}^{V}(-q_\perp^2) \right] 
\nn 
&+& \frac{q_\perp^4}{24m^4} \left[ -2 F_{1,1}^{V}(-q_\perp^2) + F_{2,1}^{V}(-q_\perp^2) \right] \,
,\\
\mathcal{Z}_1(-q_\perp^2)&=&  \frac{q_\perp^2}{6 m^2} \left[ 3 F_{1,0}^{V}(-q_\perp^2) - 2 F_{2,0}^{V}(-q_\perp^2) \right] + \frac{q_\perp^4}{24 m^4} \left[ 3 F_{1,1}^{V}(-q_\perp^2) - 2 F_{2,1}^{V}(-q_\perp^2) \right]
\,,\\
\mathcal{Z}_2(-q_\perp^2)&=& - {\frac{1}{6 }} \left[- 2 F_{1,0}^{V}(-q_\perp^2) +  F_{1,1}^{V}(-q_\perp^2)+  2 F_{2,0}^{V}(-q_\perp^2) \right]  + {\frac{q_\perp^2}{12 m^2}} \left[ F_{1,1}^{V}(-q_\perp^2) - F_{2,1}^{V}(-q_\perp^2) \right] \,,
\\
\mathcal{A}_0(-q_\perp^2)&=&  F_{1,0}^{V}(-q_\perp^2)  + \frac{q_\perp^2}{30 m^2} \left[ - 2 F_{1,0}^{V}(-q_\perp^2) + 7  F_{1,1}^{V}(-q_\perp^2)  \right] - \frac{q_\perp^4}{60 m^4}  F_{1,1}^{V}(-q_\perp^2)\, ,
\\
\mathcal{A}_1(-q_\perp^2)&=&   \frac{1}{3 }F_{2,0}^{V}(-q_\perp^2)  + \frac{q_\perp^2}{15 m^2} \left[  -F_{2,0}^{V}(-q_\perp^2) +  F_{2,1}^{V}(-q_\perp^2)  \right] - \frac{q_\perp^4}{60 m^4}  F_{2,1}^{V}(-q_\perp^2)\, , 
\\
\mathcal{A}_2(-q_\perp^2)&=& \frac{q_\perp^2}{6  m^2}F_{1,0}^{V}(-q_\perp^2) + \frac{q_\perp^4}{24 m^4}  F_{1,1}^{V}(-q_\perp^2)\, ,
\\
\mathcal{A}_3(-q_\perp^2)&=& -{\frac{1}{6}} F_{1,1}^{V}(-q_\perp^2) \, , 
\\
\mathcal{A}_4(-q_\perp^2)&=& \frac{q_\perp^2}{6  m^2}F_{2,0}^{V}(-q_\perp^2)  + \frac{q_\perp^4}{24 m^4}  F_{2,1}^{V}(-q_\perp^2) \, ,
\\
\mathcal{A}_5(-q_\perp^2)&=& - {\frac{1}{6}}F_{2,1}^{V}(-q_\perp^2)\, .
\end{eqnarray}
Coefficient functions for $ t^{00}_{\phi}(s',s,{\bf r}) $ and $ t^{ij}_{\phi,0}(s',s,{\bf r}) $: 
\begin{eqnarray}
\mathcal{E}_0\left(q_{\perp}^{2}\right)&=& F_{1,0}(-q_\perp^2) - \frac{2}{3} F_{6,0}(-q_\perp^2) +\frac{q_\perp^2}{3m^2} \left[ -F_{1,0}(-q_\perp^2) + \frac{1}{2} F_{1,1}(-q_\perp^2) + F_{4,0}(-q_\perp^2) +2 F_{5,0}(-q_\perp^2) \right] \nn
&&+  \frac{q_\perp^4}{12m^4} \left[ -F_{1,1}(-q_\perp^2) + F_{4,1}(-q_\perp^2) \right],\\
\mathcal{E}_1\left(q_{\perp}^{2}\right)&=&  \frac{2}{3} F_{6,0}(-q_\perp^2) +\frac{q_\perp^2}{m^2} \left[ \frac{1}{2} F_{1,0}(-q_\perp^2) -\frac{2}{3} F_{4,0}(-q_\perp^2) - \frac23 F_{5,0}(-q_\perp^2) \right] 
\nn
&+& \frac{q_\perp^4}{m^4} \left[ \frac18 F_{1,1}(-q_\perp^2) - \frac16 F_{4,1}(-q_\perp^2) \right],
\\
\mathcal{E}_2\left(q_{\perp}^{2}\right)&=& {\frac{1}{3}} \left[ F_{1,0}(-q_\perp^2) -\frac12 F_{1,1}(-q_\perp^2) -2 F_{4,0}(-q_\perp^2) \right] + {\frac{q_\perp^2}{12m^2}} \left[ F_{1,1}(-q_\perp^2) - 2 F_{4,1}(-q_\perp^2)\right], 
\end{eqnarray}
\begin{eqnarray}
{a}_{1}\left(r\right) &=& \int\frac{d^{3}q}{(2\pi)^{3}}\,e^{-i{\bf q}\cdot{\bf r}} \int d^{2}\hat{n} \, \frac{q_{\perp}^{2}}{2q^{2}} \bigg\{
F_{1,0}(-q_\perp^2) - \frac{2}{3} F_{6,0}(-q_\perp^2) 
\nn
&-& \frac{q_\perp^2}{3m^2} \left[ F_{1,0}(-q_\perp^2) - \frac{1}{2} F_{1,1}(-q_\perp^2) - F_{4,0}(-q_\perp^2) -2 F_{5,0}(-q_\perp^2) \right] 
-  \frac{q_\perp^4}{12m^4} \left[ F_{1,1}(-q_\perp^2) - F_{4,1}(-q_\perp^2) \right] \bigg\}, \ \
\\ 
{a}_{2}\left(r\right) &=& \int\frac{d^{3}q}{(2\pi)^{3}}\,e^{-i{\bf q}\cdot{\bf r}} \int d^{2}\hat{n} \, \frac{1}{q^{2}}\left(1-\frac{3q_{\perp}^{2}}{2q^{2}}\right) \bigg\{
F_{1,0}(-q_\perp^2) - \frac{2}{3} F_{6,0}(-q_\perp^2) 
\nn
&-& \frac{q_\perp^2}{3m^2} \left[ F_{1,0}(-q_\perp^2) - \frac{1}{2} F_{1,1}(-q_\perp^2) - F_{4,0}(-q_\perp^2) -2 F_{5,0}(-q_\perp^2) \right] 
-  \frac{q_\perp^4}{12m^4} \left[ F_{1,1}(-q_\perp^2) - F_{4,1}(-q_\perp^2) \right] \bigg\}, \ \ 
\\
{a}_{3}\left(r\right)&=& \int\frac{d^{3}q}{(2\pi)^{3}}\,e^{-i{\bf q}\cdot{\bf r}} \int d^{2}\hat{n}\; \frac{q_{\perp}^{4}}{4q^{4}}
\bigg\{\frac{2}{3}F_{6,0}\left(-q_{\perp}^{2}\right)+\frac{q^{2}}{6m^{2}}\left[2F_{1,0}\left(-q_{\perp}^{2}\right)-F_{1,1}\left(-q_{\perp}^{2}\right)-4F_{4,0}\left(-q_{\perp}^{2}\right)\right]
\nn
&+&\frac{q_{\perp}^{2}}{6m^{2}}\left[F_{1,0}\left(-q_{\perp}^{2}\right)+F_{1,1}\left(-q_{\perp}^{2}\right)-4F_{5,0}\left(-q_{\perp}^{2}\right)\right]+\frac{q^{2}q_{\perp}^{2}}{12m^{4}}\left[F_{1,1}\left(-q_{\perp}^{2}\right)-2F_{4,1}\left(-q_{\perp}^{2}\right)\right]
\nn
&+&\frac{q_{\perp}^{4}}{24m^{4}}F_{1,1}\left(-q_{\perp}^{2}\right)\bigg\} ,
\\
{a}_{4}\left(r\right)&=&   \int\frac{d^{3}q}{(2\pi)^{3}}\,e^{-i{\bf q}\cdot{\bf r}} \int d^{2}\hat{n}\; 
\frac{1}{3q^{4}}\bigg\{2F_{6,0}\left(-q_{\perp}^{2}\right)-\frac{10q_{\perp}^{2}}{q^{2}}F_{6,0}\left(-q_{\perp}^{2}\right)+\frac{35q_{\perp}^{4}}{4q^{4}}F_{6,0}\left(-q_{\perp}^{2}\right)
\nn
&+&\frac{q_{\perp}^{2}}{4m^{2}}\left[4F_{1,0}\left(-q_{\perp}^{2}\right)+F_{1,1}\left(-q_{\perp}^{2}\right)-4F_{4,0}\left(-q_{\perp}^{2}\right)-8F_{5,0}\left(-q_{\perp}^{2}\right)\right]
\nn
&-&
\frac{5q_{\perp}^{4}}{8m^{2}q^{2}}\left[5F_{1,0}\left(-q_{\perp}^{2}\right)+\frac{7}{2}F_{1,1}\left(-q_{\perp}^{2}\right)-2F_{4,0}\left(-q_{\perp}^{2}\right)-16F_{5,0}\left(-q_{\perp}^{2}\right)\right]
\nn
&+&
\frac{q_{\perp}^{4}}{4m^{4}}\left[F_{1,1}\left(-q_{\perp}^{2}\right)-F_{4,1}\left(-q_{\perp}^{2}\right)\right]+\frac{35q_{\perp}^{6}}{16m^{2}q^{4}}\left[F_{1,0}\left(-q_{\perp}^{2}\right)+F_{1,1}\left(-q_{\perp}^{2}\right)-4F_{5,0}\left(-q_{\perp}^{2}\right)\right]
\nn
&+&
\frac{5q_{\perp}^{6}}{32m^{4}q^{2}}\left[-5F_{1,1}\left(-q_{\perp}^{2}\right)+2F_{4,1}\left(-q_{\perp}^{2}\right)\right] +\frac{35q_{\perp}^{8}}{64m^{4}q^{4}}F_{1,1}\left(-q_{\perp}^{2}\right)\bigg\} ,
\\
{a}_{5}\left(r\right) &=&   \int\frac{d^{3}q}{(2\pi)^{3}}\,e^{-i{\bf q}\cdot{\bf r}} \int d^{2}\hat{n}\;  
\frac{q_{\perp}^{2}}{12q^{4}}\bigg\{4F_{6,0}\left(-q_{\perp}^{2}\right)-\frac{5q_{\perp}^{2}}{q^{2}}F_{6,0}\left(-q_{\perp}^{2}\right)
\nn
&+&
\frac{q_{\perp}^{2}}{2m^{2}}\left[5F_{1,0}\left(-q_{\perp}^{2}\right)+\frac{1}{2}F_{1,1}\left(-q_{\perp}^{2}\right)-6F_{4,0}\left(-q_{\perp}^{2}\right)-8F_{5,0}\left(-q_{\perp}^{2}\right)\right]
\nn
&-&
\frac{5q_{\perp}^{4}}{4m^{2}q^{2}}\left[F_{1,0}\left(-q_{\perp}^{2}\right)+F_{1,1}\left(-q_{\perp}^{2}\right)-4F_{5,0}\left(-q_{\perp}^{2}\right)\right]+\frac{q_{\perp}^{4}}{8m^{4}}\left[5F_{1,1}\left(-q_{\perp}^{2}\right)-6F_{4,1}\left(-q_{\perp}^{2}\right)\right]
\nn
&-&
\frac{5q_{\perp}^{6}}{16m^{4}q^{2}}F_{1,1}\left(-q_{\perp}^{2}\right)\bigg\}\,,
\\
{a}_{6}\left(r\right) &=&   \int\frac{d^{3}q}{(2\pi)^{3}}\,e^{-i{\bf q}\cdot{\bf r}} \int d^{2}\hat{n}\; 
\frac{q_{\perp}^{2}}{3q^{4}}\bigg\{2F_{6,0}\left(-q_{\perp}^{2}\right)-\frac{5q_{\perp}^{2}}{2q^{2}}F_{6,0}\left(-q_{\perp}^{2}\right)
\nn
&+&
\frac{q_{\perp}^{2}}{8m^{2}}\left[2F_{1,0}\left(-q_{\perp}^{2}\right)+5F_{1,1}\left(-q_{\perp}^{2}\right)+4F_{4,0}\left(-q_{\perp}^{2}\right)-16F_{5,0}\left(-q_{\perp}^{2}\right)\right]
\nn
&-&
\frac{5q_{\perp}^{4}}{8m^{2}q^{2}}\left[F_{1,0}\left(-q_{\perp}^{2}\right)+F_{1,1}\left(-q_{\perp}^{2}\right)-4F_{5,0}\left(-q_{\perp}^{2}\right)\right]+\frac{q_{\perp}^{4}}{16m^{4}}\left[F_{1,1}\left(-q_{\perp}^{2}\right)+2F_{4,1}\left(-q_{\perp}^{2}\right)\right]
\nn
&-&
\frac{5q_{\perp}^{6}}{32m^{4}q^{2}}F_{1,1}\left(-q_{\perp}^{2}\right)\bigg\}.
\end{eqnarray}
Coefficient functions for $ t^{ij}_{\phi,2}(s',s,{\bf r})$:
\begin{eqnarray}
\mathcal{W}_0\left(q_{\perp}^{2}\right) &= & F_{2,0}(-q_\perp^2) + \frac{q_\perp^2}{6m^2} \left[ -2 F_{2,0}(-q_\perp^2) + F_{2,1}(-q_\perp^2) \right] - \frac{q_\perp^4}{12m^4} F_{2,1}(-q_\perp^2) ,
\\
\mathcal{W}_1\left(q_{\perp}^{2}\right) &= & \frac{q_\perp^2}{2m^2} F_{2,0}(-q_\perp^2) +  \frac{q_\perp^4}{8m^4} F_{2,1}(-q_\perp^2) ,
\\
\mathcal{W}_2\left(q_{\perp}^{2}\right) &= & {\frac{1}{6}} \left[ 2 F_{2,0}(-q_\perp^2) - F_{2,1}(-q_\perp^2)  \right] + {\frac{q_\perp^2}{12m^2}} F_{2,1}(-q_\perp^2) ,
\\
\mathcal{U}_0\left(q_{\perp}^{2}\right) &= & F_{3,0}(-q_\perp^2) + \frac{q_\perp^2}{6m^2} \left[ -2 F_{3,0}(-q_\perp^2) + F_{3,1}(-q_\perp^2) \right] - \frac{q_\perp^4}{12m^4} F_{3,1}(-q_\perp^2) ,
\\
\mathcal{U}_1\left(q_{\perp}^{2}\right) &= & \frac{q_\perp^2}{2m^2} F_{3,0}(-q_\perp^2) +  \frac{q_\perp^4}{8m^4} F_{3,1}(-q_\perp^2) ,
\\
\mathcal{U}_2\left(q_{\perp}^{2}\right) &= & {\frac{1}{6}} \left[ 2 F_{3,0}(-q_\perp^2) - F_{3,1}(-q_\perp^2)  \right] + {\frac{q_\perp^2}{12m^2}} F_{3,1}(-q_\perp^2) ,
\\
{v}_{0}\left(r\right)&=&\int\frac{d^{3}q}{\left(2\pi\right)^{3}}e^{-i\vec{q}\cdot\vec{r}}\int d^{2}\hat{n} \Biggl\{  -2 F_{3,0}(-q_\perp^2) + \frac{q_\perp^2}{3 m^2} \left[ -\frac{3}{2} F_{2,0}(-q_\perp^2) +2 F_{3,0}(-q_\perp^2) -F_{3,1}(-q_\perp^2)\right] \nn &+& \frac{q_\perp^4}{6 m^4} \left[ F_{2,0}(-q_\perp^2) - \frac{1}{2} F_{2,1}(-q_\perp^2) +F_{3,1}(-q_\perp^2)\right] + \frac{q_\perp^6}{24 m^6}  F_{2,1}(-q_\perp^2) \Biggl\},
 \\
{v}_{1} \left(r\right)&=&\int\frac{d^{2}\hat{n}d^{3}q}{\left(2\pi\right)^{3}}e^{-i\vec{q}\cdot\vec{r}}\frac{q_{\perp}^{2}}{2m^{2}q^{2}}\Bigg\{\left(\frac{2}{3}-\frac{q_{\perp}^{2}}{2q^{2}}\right)\left(-\frac{q_{\perp}^{2}}{2m^{2}}F_{2,0}(-q_{\perp}^{2})-2F_{3,0}(-q_{\perp}^{2})\right) ,
\nonumber\\
&&+\frac{1}{2}\left(\frac{1}{3}+\frac{q_{\perp}^{2}}{3m^{2}}-\frac{q_{\perp}^{2}}{q^{2}}-\frac{q_{\perp}^{4}}{4m^{2}q^{2}}\right)\left(-\frac{q_{\perp}^{2}}{2m^{2}}F_{2,1}(-q_{\perp}^{2})-2F_{3,1}(-q_{\perp}^{2})\right)\Bigg\} \,.
\end{eqnarray}
 Coefficient functions for $ t^{0i}_{\phi}(s',s,{\bf r}) $:
  \begin{eqnarray}
\mathcal{C}_0\left(q_{\perp}^{2}\right) &=&  F_{1,0}(-q_\perp^2) -\frac13 F_{4,0}(-q_\perp^2) -\frac{4}{15} F_{6,0}(-q_\perp^2) 
\nn
&& +\frac{q_\perp^2}{15m^2}  \left[ - F_{1,0}(-q_\perp^2) + \frac72 F_{1,1}(-q_\perp^2) + F_{4,0}(-q_\perp^2) - F_{4,1}(-q_\perp^2) + 4 F_{5,0}(-q_\perp^2) \right] 
\nn
&& + \frac{q_\perp^4}{60m^4} \left[ -F_{1,1}(-q_\perp^2) +F_{4,0}(-q_\perp^2)  \right],
\\
\mathcal{C}_1\left(q_{\perp}^{2}\right) &=&  \frac13 F_{4,0}(-q_\perp^2) + \frac{q_\perp^2}{15m^2} \left[ -F_{4,0}(-q_\perp^2)  + F_{4,1}(-q_\perp^2) \right] -  \frac{q_\perp^4}{60m^4}  F_{4,1}(-q_\perp^2)  ,
\\
\mathcal{C}_2\left(q_{\perp}^{2}\right) &=&  \frac23 F_{6,0}(-q_\perp^2) + \frac{q_\perp^2}{6m^2} \left[ F_{1,0}(-q_\perp^2) - F_{4,0}(-q_\perp^2) - 4 F_{5,0}(-q_\perp^2) \right] \nn
&&+ \frac{q_\perp^4}{24m^4}  \left[ F_{1,1}(-q_\perp^2) - F_{4,1}(-q_\perp^2) \right],
\\
\mathcal{C}_3\left(q_{\perp}^{2}\right) &=&  {\frac{1}{6}}  \left[ - F_{1,1}(-q_\perp^2) + F_{4,1}(-q_\perp^2) \right] ,
\\
\mathcal{C}_4\left(q_{\perp}^{2}\right) &=&  \frac{q_\perp^2}{6m^2} F_{4,0}(-q_\perp^2) +  \frac{q_\perp^4}{24m^4} F_{4,1}(-q_\perp^2) ,
\\
\mathcal{C}_5\left(q_{\perp}^{2}\right) &=&  -{\frac{1}{6}} F_{4,1}(-q_\perp^2) .
\end{eqnarray}


\begin{references}

\bibitem{Hofstadter:1958}
R.~Hofstadter, F.~Bumiller, and M.~R.~Yearian,
Rev. Mod. Phys. {\bf 30}, 482 (1958).

\bibitem{Ernst:1960zza}
F.~J.~Ernst, R.~G.~Sachs and K.~C.~Wali,
Phys. Rev. \textbf{119}, 1105-1114 (1960).

\bibitem{Sachs:1962zzc}
R.~G.~Sachs,
Phys. Rev. \textbf{126}, 2256-2260 (1962).

\bibitem{Polyakov:2002wz}
M.~V.~Polyakov and A.~G.~Shuvaev,
[arXiv:hep-ph/0207153 [hep-ph]].

\bibitem{Polyakov:2002yz} 
  M.~V.~Polyakov,
  Phys.\ Lett.\ B {\bf 555}, 57 (2003). 

\bibitem{Polyakov:2018zvc}
M.~V.~Polyakov and P.~Schweitzer,
Int.\ J.\ Mod.\ Phys.\ A \textbf{33} (2018) no.26, 1830025. 

\bibitem{Burkardt:2000za}
M.~Burkardt,
Phys. Rev. D \textbf{62} (2000), 071503(R),
[erratum: Phys. Rev. D \textbf{66} (2002), 119903(E)].

\bibitem{Miller:2007uy}
G.~A.~Miller,
Phys. Rev. Lett. \textbf{99}, 112001 (2007).

\bibitem{Miller:2009qu}
G.~A.~Miller,
Phys. Rev. C \textbf{79}, 055204 (2009).

\bibitem{Miller:2010nz}
G.~A.~Miller,
Ann. Rev. Nucl. Part. Sci. \textbf{60} (2010), 1-25.

\bibitem{Jaffe:2020ebz}
R.~L.~Jaffe,
Phys. Rev. D \textbf{103} (2021) no.1, 016017.

\bibitem{Miller:2018ybm}
G.~A.~Miller,
Phys. Rev. C \textbf{99}, no.3, 035202 (2019).


\bibitem{Freese:2021czn}
A.~Freese and G.~A.~Miller,
Phys. Rev. D \textbf{103}, 094023 (2021).


\bibitem{Lorce:2018egm}
C.~Lorc\'e, H.~Moutarde and A.~P.~Trawi\'nski,
Eur. Phys. J. C \textbf{79}, no.1, 89 (2019),
[arXiv:1810.09837 [hep-ph]].


\bibitem{Lorce:2020onh}
C.~Lorc\'e,
Phys. Rev. Lett. \textbf{125}, no.23, 232002 (2020),
[arXiv:2007.05318 [hep-ph]].

\bibitem{Lorce:2022cle}
C.~Lorc\'e, P.~Schweitzer and K.~Tezgin,
[arXiv:2202.01192 [hep-ph]].


\bibitem{Chen:2022smg}
Y.~Chen and C.~Lorc\'e,
[arXiv:2210.02908 [hep-ph]].


\bibitem{Guo:2021aik}
Y.~Guo, X.~Ji and K.~Shiells,
Nucl. Phys. B \textbf{969}, 115440 (2021),
[arXiv:2101.05243 [hep-ph]].


\bibitem{Panteleeva:2021iip}
J.~Y.~Panteleeva and M.~V.~Polyakov,
Phys. Rev. D \textbf{104} (2021) no.1, 014008,
[arXiv:2102.10902 [hep-ph]].


\bibitem{Panteleeva:2022khw}
J.~Y.~Panteleeva, E.~Epelbaum, J.~Gegelia and U.-G.~Mei\ss{}ner,
Phys. Rev. D \textbf{106}, no.5, 056019 (2022),
[arXiv:2205.15061 [hep-ph]].

\bibitem{Epelbaum:2022fjc}
E.~Epelbaum, J.~Gegelia, N.~Lange, U.~G.~Mei\ss{}ner and M.~V.~Polyakov,
Phys. Rev. Lett. \textbf{129} (2022) no.1, 012001.


\bibitem{Kim:2021kum}
J.~Y.~Kim and H.~C.~Kim,
Phys. Rev. D \textbf{104} (2021) no.7, 074003
[arXiv:2106.10986 [hep-ph]].

\bibitem{Kim:2021jjf}
J.~Y.~Kim and H.~C.~Kim,
Phys. Rev. D \textbf{104} (2021) no.7, 074019
[arXiv:2105.10279 [hep-ph]].

\bibitem{Kim:2022bia}
J.~Y.~Kim,
Phys. Rev. D \textbf{106} (2022) no.1, 014022
[arXiv:2204.08248 [hep-ph]].

\bibitem{Kim:2022wkc}
J.~Y.~Kim, B.~D.~Sun, D.~Fu and H.~C.~Kim,
[arXiv:2208.01240 [hep-ph]].

\bibitem{Freese:2021mzg}
A.~Freese and G.~A.~Miller,
Phys. Rev. D \textbf{105}, no.1, 014003 (2022),
[arXiv:2108.03301 [hep-ph]].


\bibitem{Freese:2022fat}
A.~Freese and G.~A.~Miller,
[arXiv:2210.03807 [hep-ph]].



\bibitem{Carlson:2022eps}
C.~E.~Carlson,
[arXiv:2208.00826 [hep-ph]].

\bibitem{Fleming:1974af}
G.~N.~Fleming,  
Physical Reality \& Math. Descrip., 357 (1974). 


\bibitem{Panteleeva:2022uii}
J.~Y.~Panteleeva, E.~Epelbaum, J.~Gegelia and U.-G.~Mei\ss{}ner,
[arXiv:2211.09596 [hep-ph]].

\bibitem{Pascalutsa:2006up}
V.~Pascalutsa, M.~Vanderhaeghen and S.~N.~Yang,
Phys. Rept. \textbf{437} (2007), 125-232
[arXiv:hep-ph/0609004 [hep-ph]].

\bibitem{Cotogno:2019vjb}
S.~Cotogno, C.~Lorc\'e, P.~Lowdon and M.~Morales,
Phys. Rev. D \textbf{101}, no.5, 056016 (2020),
[arXiv:1912.08749 [hep-ph]].

\bibitem{Gegelia:1994zz}
J.~Gegelia, G.~S.~Japaridze and K.~S.~Turashvili,
Theor. Math. Phys. \textbf{101}, 1313-1319 (1994).

\bibitem{Kim:2020lrs}
J.~Y.~Kim and B.~D.~Sun,
Eur. Phys. J. C \textbf{81} (2021) no.1, 85,
[arXiv:2011.00292 [hep-ph]].

\bibitem{Polyakov:2019lbq}
M.~V.~Polyakov and B.~D.~Sun,
Phys. Rev. D \textbf{100} (2019) no.3, 036003,
[arXiv:1903.02738 [hep-ph]].

\bibitem{Panteleeva:2020ejw}
J.~Y.~Panteleeva and M.~V.~Polyakov,
Phys. Lett. B \textbf{809} (2020), 135707,
[arXiv:2004.02912 [hep-ph]].

\bibitem{Alharazin:2022wjj}
H.~Alharazin, E.~Epelbaum, J.~Gegelia, U.-G.~Mei\ss{}ner and B.~D.~Sun,
Eur. Phys. J. C \textbf{82} (2022) no.10, 907,
[arXiv:2209.01233 [hep-ph]].

\end{references}
\end{document}